# Transformation-optics macroscopic visible-light cloaking beyond two dimensions

Chia-Wei Chu,[1,2,†] Xiaomin Zhai,[1,2,†] Chih Jie Lee,[2,3] Yubo Duan,[4]
Din Ping Tsai,[2,3,5*] Baile Zhang,[6,7*] and Yuan Luo[1,2,8*]

[1]Center for Optoelectronic Medicine, National Taiwan University, Taipei, Taiwan, 10051

[2]Molecular Imaging Center, Optical Imaging Core Laboratory, National Taiwan University, Taipei, Taiwan, 10617

[3]Department of Physics, National Taiwan University, Taipei, Taiwan, 10617

[4]Department of Bioengineering, Faculty of Engineering, National University of Singapore, Singapore 117574

[5]Research Center for Applied Science, Academia Sinica, Taipei, Taiwan, 11529

[6]Division of Physics and Applied Physics, School of Physical and Mathematical Sciences, Nanyang Technological University, Singapore, 637371

[7]Centre for Disruptive Photonic Technologies, Nanyang Technological University, Singapore 637371

[8]Department of Optics and Photonics, National Central University, Taoyuan, Taiwan, 32001

†*Authors contributed equally to this work*

**To whom correspondence should be addressed;*

*Email: dptsai@phys.ntu.edu.tw; blzhang@ntu.edu.sg; yuanluo@ntu.edu.tw*

**Transformation optics [1, 2], a recent geometrical design strategy of controlling light by combining Maxwell's principles of electromagnetism with Einstein's general relativity, promises without precedent an invisibility cloaking device that can render a macroscopic object invisible in three dimensions [2]. However, most previous proof-of-concept transformation-optics cloaking devices focused predominantly on two dimensions [3-13], whereas detection of a macroscopic object along its third dimension was always unfailing. Here, we report the first experimental demonstration of transformation-optics macroscopic visible-light cloaking beyond two dimensions. This almost-three-dimensional cloak exhibits three-dimensional (3D) invisibility for illumination near its center (i.e. with a limited field of view), and its ideal wide-angle invisibility performance is preserved in multiple two-dimensional (2D) planes intersecting in the 3D space. Both light ray trajectories and optical path lengths have been verified experimentally at the macroscopic scale, which provides unique evidence on the geometrical nature of transformation optics.**

The geometrical perception of optical space from the viewpoint of light is manifested in both the light ray trajectory and the optical path length, whose significance has been recognized back to the age of Fermat. However, it is not until recently that geometrical schemes are systematically applied in designing optical devices by artificially creating a virtual optical space through a coordinate transformation, whose approach is now widely called transformation optics [1, 2, 14-16]. A unique promise made by transformation optics is an invisibility cloaking device that can render a macroscopic 3D object invisible, as firstly proposed by J. Pendry *et al*. [2], which vividly illustrates the geometrical nature of transformation optics that both light ray trajectory and optical path length are controlled in the 3D space. The first transformation-optics invisibility cloak was implemented in a 2D plane to hide an about one-wavelength large object in a narrow microwave frequency band [3]. To extend the bandwidth, a dielectric isotropic carpet cloak designed from 2D quasi-conformal mapping [4] was proposed that can hide an object sitting on a flat ground plane, and was subsequently realized in both microwave [5] and infrared frequencies [6, 7]. Attempt to extend dimensions from 2D to 3D has been made by rotating the original 2D carpet cloak design around its central axis [8] or extending it along the third dimension [9]. However, the original 2D carpet cloak design intrinsically possesses a lateral shift of ray trajectory that is comparable to the height of the hidden object [17], and thus did not manage to hide a macroscopic object. On the other hand, a modified carpet cloak model that preserved anisotropy [12,13,18,19] successfully realized macroscopic invisibility for visible light in a 2D geometry. Yet how to extend further the dimensions of a macroscopic invisibility cloak, and whether the geometrical transformation-optics strategy applies in a macroscopic 3D space, still remain unknown.

Here we adopt a simplified almost-3D cloak to demonstrate that transformation optics can physically apply beyond two dimensions at the macroscopic scale with both light ray trajectory and optical path length precisely controlled. A perfect 3D cloak in theory that is invisible for all viewing angles can be designed by squeezing the optical space inside a truncated multifaceted cone, as illustrated in Fig. 1a, which consists of one octagonal cylinder located at the center, and eight triangular prisms and eight triangular pyramids fanned out on the periphery. (See Supplementary Information for a detailed mathematical design.) This cloak squeezes optical space by lifting the bottom of the octagonal cylinder to a height of $h$, when the original height of the octagonal cylinder is $H$. A perfectly hidden physical space, as indicated in yellow in Fig. 1a, can be created under the squeezed optical space. This ideal 3D cloak requires anisotropic magnetism [2], being extremely difficult in reality. To simplify the implementation, we first fix the light polarization to be the one with magnetic field parallel to the ground plane. We then preserve the 3D invisibility performance of the cloak within a limited field of view (i.e. when the light is incident on the octagonal cylinder), and preserve the wide-angle invisibility performance of the cloak in multiple 2D planes intersecting in the 3D space (i.e. when the light is incident on sides of the triangular prisms along axial directions). An illustration of the simplified cloak as well as its invisible illumination region and invisible planes of incidence is plotted in Fig. 1b. We completely remove triangular pyramids on the periphery as unimportant for demonstration. The materials in the octagonal cylinder (region I) and triangular prisms (region II) have two orthogonal principal refractive indices $n_{i,1}$ and $n_{i,2}$, where $i$ = I, II. For the parameters chosen in experiment (e.g. $H$=10.64 mm, $h$=0.64 mm), transformation optics requires $n_{I,1}^2=1.649^2$, $n_{I,2}^2=1.529^2$, $n_{II,1}^2=1.638^2$, and $n_{II,2}^2=1.54^2$, when the ambient environment possesses a refractive index $n$=1.54. We adopt a natural negative uniaxial crystal BaB$_2$O$_4$ (α-BBO; $n_o$=~1.64, $n_e$= ~1.53) to implement this design. All nice pieces of α-BBO crystal are cemented together with an optical liquid photopolymer (Norland Optical Adhesive 61), which cures after exposure to ultraviolet light.

We then proceed to demonstrate the almost-3D invisibility performance of the simplified cloak. The first step is to verify ray trajectories in the physical space as predicted by transformation optics. The experimental setup is illustrated in Fig. 2a. The cloak sitting on a mirror which served as the flat ground plane was immersed in a glass tank filled with a colorless laser liquid (Cargille Labs, Code 1074; $n$=1.54 measured at wavelength 589 nm). A diffractive object (a 0.5 mm×10 mm×10 mm reflective grating with groove period of 4.8 μm) was selected as the object to be hidden. A transmission pattern of two circular holes with the same diameter of 1.5 mm was illuminated by a continuous-wave laser diode at a wavelength of 532 nm (green) polarized with magnetic field parallel to the ground plane. The light transmitted through the left circular hole (Ray 1) went through the cloak with the hidden diffractive object underneath and was reflected at the bottom of the cloak, while the light transmitted through the right circular hole (Ray 2) was reflected on the mirror surface directly. A digital camera (Canon EOS 650D) was used to capture light spots projected on a screen about 5 cm away from the cloak. Figures 2b and c show images for a fixed azimuthal viewing angle φ = 0° and a fixed incidence angle θ = 25°. The uncloaked diffractive object generated multiple light spots on the screen because of grating diffraction of Ray 1, while the flat mirror surface produced only one light spot because of the mirror reflection of Ray 2. When the diffractive object was hidden by the cloak as in Fig. 2c, the original multiple light spots on the screen turned to be only one located at the same altitude of the light spot reflected from the flat mirror surface, as if the diffractive object were not there. This demonstration on recovering ray trajectories as predicted by transformation optics is similar to previous experiments [6,7,9,12,13].

To further prove the beyond-2D invisibility performance in a broad bandwidth, measurements were performed at different azimuthal viewing angles with other wavelengths at 473 nm (blue) and 636 nm (red). With the fixed incidence angle θ = 25°, output profiles of light spots on the screen for the azimuthal viewing angles of φ = 0° and 45° with 473 nm wavelength, and those for φ = 90° and 135° with 636 nm wavelength, are shown in Fig. 3a-b and c-d, respectively. Light spots reflected from the diffractive object without the cloak changed their orientation at different azimuthal viewing angles. Intensity of these light spots changed with wavelength because of the intrinsic dispersion of diffractive object. In Figs. 3e-h, when the diffractive object was cloaked, its reflection exhibited only one light spot at the same altitude as the one reflected from the flat mirror surface, which demonstrates clearly the cloaking phenomenon at different azimuthal viewing angles. More results at different incidence angles are provided in Supplementary Information.

The second step of verifying the almost-3D invisibility performance is to measure the optical path length, which links to the distance for which a light ray passes in the virtual space. For the purpose of high accuracy without phase unwrapping, a low-coherence interferometer with a Ti: sapphire mode-locked femtosecond laser (Spectra Physics, Mountain View, CA) was adopted for measurement. To ensure high accuracy, we put the setup in air rather than in the laser liquid, whose validity has been justified in Ref. 20. As shown in Fig. 4a, laser pulses with magnetic field parallel to the ground plane passed through a half-wave plate and a linear polarizer in order to fix the polarization with high accuracy. Mirror 1 (M1) defined the reference path in air, whereas Mirror 2 (M2) reflected back the light going through the cloak and defined the signal path. When an interference pattern was observed at the exit arm of the interferometer, precise optical path lengths of the reference path and the signal path were matched within the coherence length of the laser source. The coherence length of the laser source (central wavelength of ~800 nm; spectral bandwidth of ~20 nm) was calculated to be 14 μm [20, 21], which we take as the nominal axial accuracy in the optical path length measurement. By comparing the locations of the reference mirror M1 in two measurements—one with the laser beam reflected from M2 going through the cloak; the other when there was only the flat mirror without the cloak—the difference in optical path lengths can be determined, and the height of the hidden space can be calculated (See Supplementary Information). For simplicity, we only consider illumination in the 3D invisible illumination cone, or when the illumination is incident on the top of the octagonal cylinder. The measurement was repeated for different incidence angles (θ) between 65$^o$ and 90$^o$ at various azimuthal viewing angles. Results of the calculated height of the hidden space at azimuthal viewing angles of φ = 0$^o$ and 45$^o$ are shown in Fig. 4b, being consistent with the real height. Measurements with other azimuthal viewing angles showed similar results.

These above results demonstrate almost-3D cloaking of a macroscopic object with both ray trajectories and optical path lengths fully controlled, and provide direct and solid evidence that transformation optics strategy can apply beyond two dimensions at the macroscale. Although there is still a long way to go to realize completely 3D invisibility, our study has made a significant step forward toward 3D optical devices using transformation optics.

## Acknowledgements

We acknowledge financial support from Taiwan National Science Council (100-2218-E-002-026-MY3 and 102-2745-M-002-005-ASP / NSC102 - 2811 - M - 002 - 084), Taiwan National Health Research Institutes (EX102-10220EC), National Taiwan University (102R7832), Nanyang Technological University, Singapore Ministry of Education under Grant No. Tier 1 RG27/12 and Grant No. MOE2011-T3-1-005. We thank Jer-Liang Andrew Yew, Yu Chiang Wang for making the diffractive object.

## Author Contributions

C.W.C. and Y.L. fabricated the cloak and performed measurements. X.Z. carried out the fabrication of the diffractive object. C.W.C., X.Z., Y.D. and C.J.L. did simulation. B.Z., D.P.T., and Y.L. designed experiments, supervised the project, and wrote the manuscript. All authors discussed the results and commented on the manuscript at all stages.

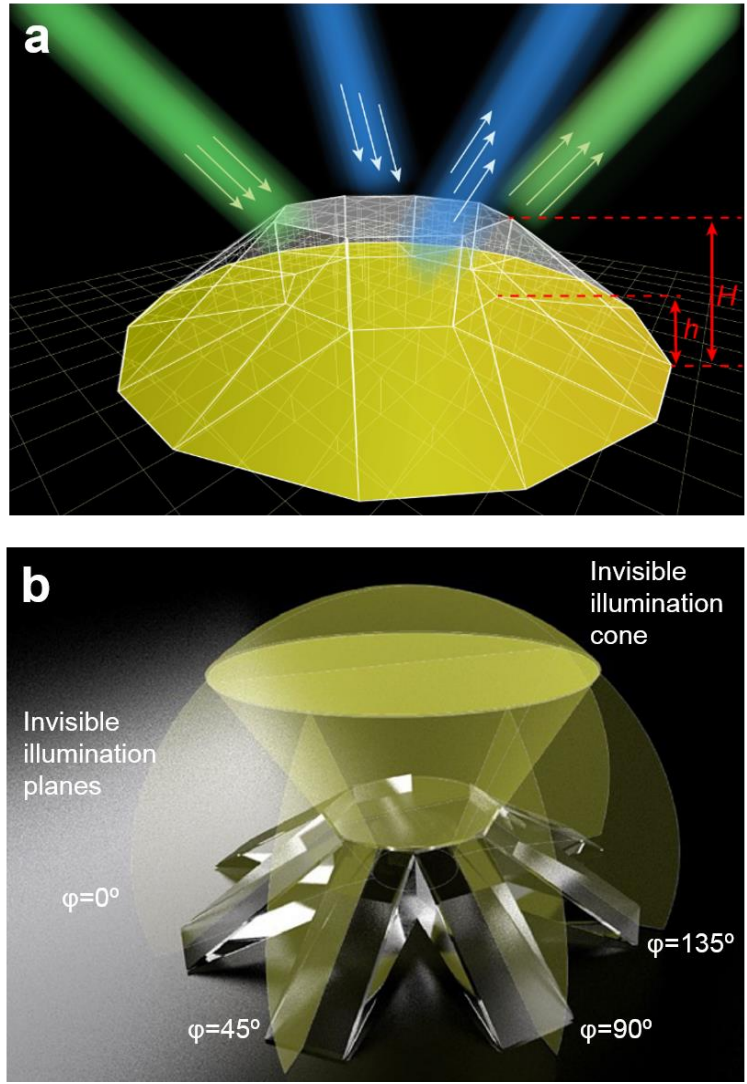

**Figure 1: Illustration of perfect three-dimension cloak and simplified almost-three-dimensional cloak. a**, Schematic of a transformation-based perfect three-dimensional cloak which consists of one octagonal cylinder located at the center, and eight triangular prisms and eight triangular pyramids fanned out on the periphery. Arbitrary incident rays will be reflected as if the object hidden under the cloak together with the cloak did not exist. **b**, Schematic of the simplified almost-three-dimensional cloak which can be implemented with α-BBO crystal. Three-dimensional invisibility is preserved within a limited field of view, or inside the invisible illumination cone. Wide-angle invisibility is preserved in multiple two-dimensional planes, or invisible illumination planes, which intersect in the three-dimensional space.

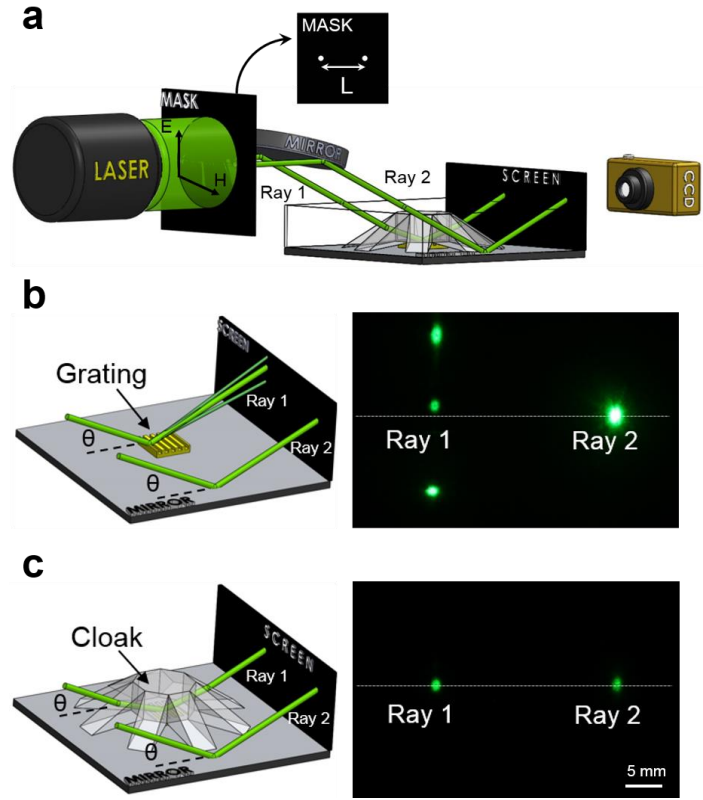

**Figure 2: Experimental setup and results for verifying ray trajectories through the almost-three-dimensional cloak. a**, The cloak that hides a diffractive object is immersed in a colorless laser liquid with refractive index 1.54. A green laser beam at the wavelength of 532 nm illuminates a two-hole mask pattern (*L*=24 mm) on the left side of the tank. Light transmitted through the left hole (Ray 1) can go through the cloak, while light through the right hole (Ray 2) will be reflected on the flat ground plane directly. A digital camera captures light spots shown on the screen behind the tank. In results of **b**-**c**, the azimuthal viewing angle is fixed at $\varphi = 0^o$ and the incidence angle is fixed at $\theta = 25^o$. **b**, When the diffractive object sits on top of the mirror surface without the cloak, its reflection generates multiple light spots, while light reflected on the flat mirror surface has only one light spot. **c**, When the cloak is hiding the diffractive object, the original multiple light spots turn to be only one at the same altitude of the light spot reflected from the mirror surface.

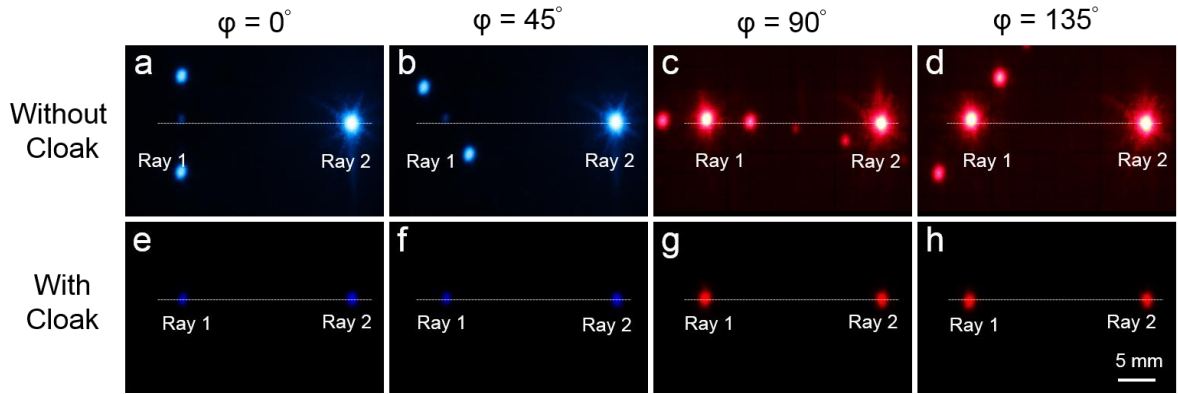

**Figure 3: Experimental results for verifying ray trajectories through the almost-three-dimensional cloak at different azimuthal viewing angles with different wavelengths. a-b**, Light spots captured on the screen when there is no cloak hiding the diffractive object at a wavelength of 473 nm (blue) along the viewing planes of φ = 0° and 45°, respectively. **c-d**, Light spots captured on the screen when there is no cloak hiding the diffractive object at a wavelength of 636 nm (red) along the viewing planes of φ = 90° and 135°, respectively. **e-h**, Light spots captured on the screen when the cloak is hiding the diffractive object at various viewing angles and wavelengths corresponding to **a-d**, respectively. In all results, the incidence angle is fixed at θ = 25°.

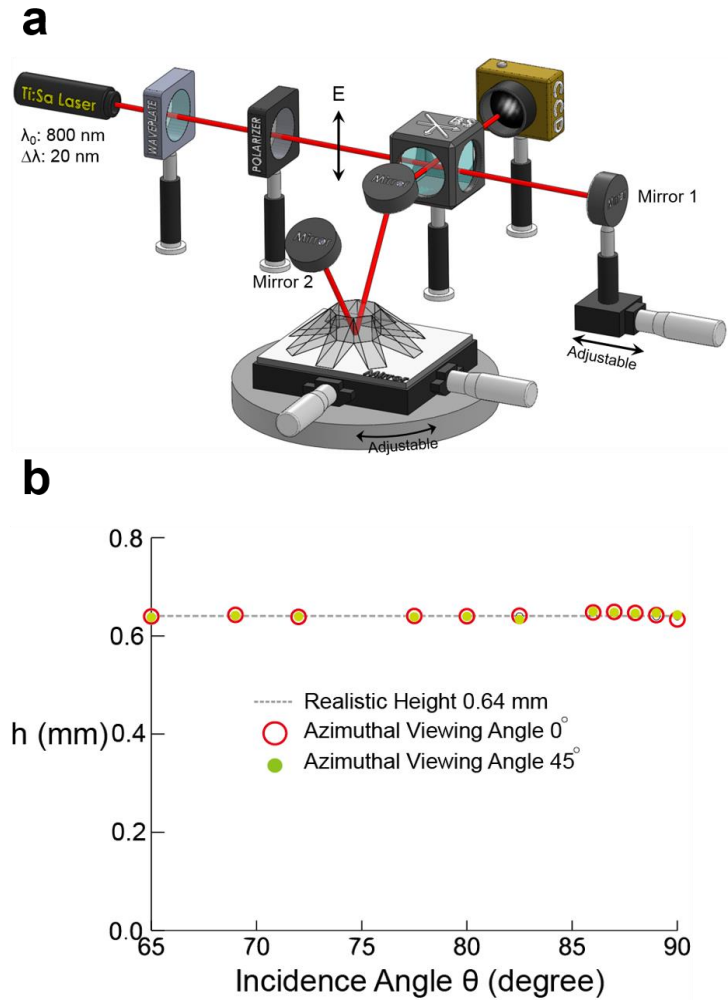

**Figure 4: Optical path length measurement setup and experimental results. a**, Experimental setup for the optical path length measurement. A low-coherence interferometer with a femtosecond laser is adopted. The cloak is illuminated with adjustable incidence angles and azimuthal viewing angles. **b**, Height of the hidden space calculated from measured optical path length at different incidence angles (θ) and azimuthal viewing planes (φ).